\documentclass[aps,prb,twocolumn,superscriptaddress,showpacs]{revtex4}
\usepackage{graphicx} 
\usepackage{dcolumn} 
\usepackage{amssymb}

\begin{document} \title{The Correlation Energy and the  Spin Susceptibility of
the Two-Valley  Two-dimensional Electron Gas.}

\author{M. Marchi} \affiliation{INFM-CNR DEMOCRITOS National
Simulation Center, Trieste, Italy} \affiliation{International School
for Advanced Studies (SISSA), via Beirut 2-4, 34014 Trieste, Italy}

\author{S.\ De Palo} \affiliation{INFM-CNR DEMOCRITOS National
Simulation Center, Trieste, Italy} \affiliation{Dipartimento di Fisica
Teorica, Universit$\grave{a}$ di Trieste, Strada Costiera 11, 34014
Trieste, Italy} \author{S.\ Moroni} \affiliation{INFM-CNR DEMOCRITOS
National Simulation Center, Trieste, Italy} \affiliation{International
School for Advanced Studies (SISSA), via Beirut 2-4, 34014 Trieste, Italy}

\author{Gaetano Senatore} \affiliation{INFM-CNR DEMOCRITOS National
Simulation Center, Trieste, Italy} \affiliation{Dipartimento di Fisica
Teorica, Universit$\grave{a}$ di Trieste, Strada Costiera 11, 34014
Trieste, Italy}

\begin{abstract}
  We find that the  spin susceptibility of a two-dimensional
  electron system with valley degeneracy does not grow critically at
  low densities, at variance with experimental results [A. Shashkin
  \emph{et al.}, Phys. Rev. Lett. \textbf{96}, 036403 (2006)].  We
  ascribe this apparent discrepancy to the weak disorder present in
  experimental samples. Our prediction is obtained from accurate
  correlation energies computed with state of-the-art diffusion Monte
  Carlo simulations and fitted with an analytical expression which also
  provides a local spin density functional for the system under
  investigation.
 
\end{abstract}


\maketitle

\section{Introduction}
The spin properties of low-dimensional electron systems in solid state
devices are of great interest in relation to spintronics and quantum
computing \cite{awscholom}, both at the fundamental level and for
technological applications, the long wavelength spin susceptibility of
the two-dimensional electron gas (2DEG) playing an important role in the control of
nuclear spins \cite{loss}. They are also believed to be intimately
related to the apparent metal-insulator transition (MIT) observed in
2D \cite{krav-rev,shash_therm,pud,gunawan,punnoose_nature}.  Indeed
the spin susceptibility $\chi_s$ of the 2DEG, measured with various
techniques \cite{krav-rev}, is consistently found to grow with respect
to its noninteracting Pauli value $\chi_0$, as the density is lowered
and the MIT approached \cite{shash_therm,pud,gunawan}. Recently,
experimental evidence has been given for a critical growth of $\chi_s$
in Si-MOSFETs at a finite density \cite{shash_therm} coincident,
within  experimental uncertainties, with the critical density for the
MIT \cite{krav-rev,punnoose_nature}. The qualitative question to which
we give an answer in this paper is whether such a {\it divergence} is
a property of the ideally clean two-valley (2V) 2DEG, the simplest
model of electrons confined in a Si-MOSFET \cite{ando}, or is due to
some other factor. It should be stressed from the outset that the
valley degree of freedom has qualitative effects on the 2DEG
properties, making the fully spin polarized fluid never stable
\cite{conti}, at variance with the one-valley (1V) 2DEG, and
importantly affects the MIT \cite{punnoose,spinvalley}.

Correlation plays a crucial role in the so-called EG, i.e.,
electrons  with a $1/r$ pair potential, moving in a
neutralizing charge background \cite{GV}. Its importance grows both
with lowering the density and the space dimensionality and tends to
quantitatively and often even qualitatively change the predictions of
simple schemes, such as the Hartree-Fock (HF) or the random-phase
approximation (RPA) \cite{GV}.  In the low-density strongly-correlated 
EG, which would be more properly called an electron liquid, the energy
balance determining the system properties is played on a very minute
scale and, to get meaningful predictions, a great accuracy such as the
one afforded by quantum Monte Carlo (QMC) methods is necessary
\cite{GV}. 

QMC simulations have provided over the years the method of
choice for microscopic studies of the 2DEG
\cite{cep89,kwon93,rapisarda,conti,varsano,attacca}, which recently
has been shown to provide a rather accurate model for electrons
confined in solid state devices \cite{depalo05}. However, no QMC
prediction is available for  $\chi_s$ in the 2V2DEG and other
theoretical  estimates, obtained either in RPA
\cite{dassarma,dassarma-ferrom} or with a classical mapping
\cite{dharmawardana-2v}, do not appear reliable \cite{note-chi}.
Here, to calculate $\chi_s$ we resort to extensive state-of-the-art
simulations of the 2V2DEG, using the diffusion Monte Carlo (DMC)
technique \cite{foulkes}.  We thus obtain for the first time the
dependence of the ground state energy on both the density and  the
spin polarization,    also improving on Ref.~\onlinecite{conti},
with the use of twist-averaged boundary conditions (TABC) \cite{tabc}
and trial wavefunctions including backflow (BF) \cite{kwon93}. 

\section{Correlation energy of the 2V2DEG}
In the 2V2DEG electrons possess an additional discrete degree of
freedom, \emph{i.e.} the valley flavor or index, which can be
conveniently described with a pseudospin.  One may
identify electrons with given spin and pseudospin indexes as
belonging to a species or {\it component}.  Accordingly, the
paramagnetic 2V2DEG is a four-component system, while both the fully
spin-polarized 2V2DEG and the paramagnetic 1V2DEG have
two components. For the sake of simplicity, we restrict here to the
symmetric case where the number of electrons and the spin polarization
are the same for both valleys \cite{splitting}. Thus, at zero
temperature, the state of the system is fully specified by the
coupling parameter $r_s=1/\sqrt{\pi\,n}\,a_B$ and the spin
polarization $\zeta=(n_{\uparrow}-n_{\downarrow})/n$, with $n$ the
total electron density, $a_B$ the Bohr radius,
$n_{\uparrow(\downarrow)}$ the density of up (down) spin electrons.
Below, Rydberg units are used throughout.

\subsection{Simulation details}
We have performed simulations with the fixed-phase (FP)
 \cite{carlson-ortiz93} DMC method, which gives the lowest upper bound
to the ground-state energy consistent with the many-body phase of a
suitably chosen, complex-valued trial function. For real trial
functions FP-DMC reduces to the standard fixed-node (FN) approximation
 \cite{foulkes}. A complex trial function allows using TABC
 \cite{tabc}, which reduce the size dependence of the kinetic energy by
one order of magnitude with respect to periodic boundary conditions
(PBC). Furthermore, since TABC do not require closed shells in
$k$-space there are no restrictions on the number of electrons per
component, so that the polarization can be changed by flipping any
number of spins, with fixed total number of electrons  \cite{zong_3d}.
Our trial function is the product of Slater plane-wave (PW)
determinants (one per component) and a Jastrow factor
 \cite{rapisarda}. BF correlations  \cite{kwon93} are included only for
$\zeta=0$ and $\zeta=1$, but with FN-DMC and in PBC. Their
contribution to the ground-state energy is then added to the PW
energies assuming a quadratic dependence on polarization as in
Ref.~\onlinecite{varsano,attacca}. The ground-state energy per particle
$E_N(r_s,\zeta)$, calculated for several values of $r_s$, $\zeta$, and
the electron number $N$, is recorded in Table \ref{fit_data} of Appendix A.

\subsection{Analytic representation}

\begin{table}[t]
  \caption{Parameters of the analytic representation (\ref{ecorr}) and
(\ref{alphaifittone}) of the correlation energy of the 2V2DEG,
determined from Eq.~(\ref{fittone}) by a least squares fit to the data
listed in Table \ref{fit_data}. The reduced chi square is
$\tilde{\chi}^2=4.82$. A $^*$ marks constrained parameters, whereas
$C_2$ is fixed to zero since it turned out to be irrelevant in the
fitting procedure. The parameters $\eta$, $\eta_z$, $\gamma$ and
$\gamma_z$ in Eq.~(\ref{fittone}) only concern the size extrapolation;
their optimal values are 0.056, 0.17, 2.03 and 0.45, respectively. }
\label{param_fit}
\begin{center}
\begin{tabular}{cccccc} \hline\hline & $i=0$ & $i=1$ & $i=2$ \\ \hline
$A_i$ & $-0.99870^*$ & $0.44570^*$ & $0.0082290$ \\ 
$B_i$ & $\frac{16}{3 \pi}(10-3\pi)^* $ & $-0.85288^* $ & $0.048979$ \\ 
$C_i$ & $0.62208$ & $-7.6202$ & $0$ 
\\ $D_i$ & $0.029726$ & $-1.6194$ & $-0.051302$ \\ 
$E_i$ & $1.6208 $ & $12.714 $ & $25.911 $ \\ 
$F_i$ & $-0.012856$ & $0^*$ & $0^*$ \\ 
$G_i$ & $0.66150$ & $19.692^*$ & $15.072^*$ \\ 
$H_i$ & $0.029765^*$ & $3.6334^*$ & $6.2343^*$ \\ \hline
$\beta$ & & 11.879 & \\ \hline\hline
\end{tabular}
\end{center}
\end{table}

Following Ref.~\onlinecite{attacca}, we determine the energy per particle
$E(r_s,\zeta)$ in the thermodynamic limit by fitting to the data
listed in Table \ref{fit_data} an analytic expression which embodies the
$r_s$ and $\zeta$ dependence as well as a Fermi-liquid--like size
correction:
\begin{eqnarray}
  \label{fittone} \nonumber E_N(r_s,\zeta)&=&E(r_s,\zeta)+ \Delta
  T_N(r_s,\zeta) 
  +\left(\eta  +\eta_z \zeta^2 \right)\frac{1}{r_s N}\\ &&+
  \left(\gamma +\gamma_z \zeta^2 \right)\frac{1}{(r_s N)^{3/2}}.
\end{eqnarray} The fitting parameters $\eta,\eta_z,\gamma$ and
$\gamma_z$ take into account potential energy finite-size effects,
while $\Delta T_N(r_s,\zeta)$ is the difference of the
non-interacting kinetic energy evaluated at finite $N$ with TABC and
in the thermodynamic limit. $E(r_s,\zeta)$ is customarily decomposed
as  sum of the non-interacting kinetic energy,
$\epsilon_0(r_s,\zeta)=(1+\zeta^2)/(2r_s^2)$, the exchange energy,
$e_{x}(r_s,\zeta)=-\frac{4}{3 \pi}\frac{1}{r_s}
[(1+\zeta)^{3/2}+(1-\zeta)^{3/2}]$, and the unknown correlation energy
$e_c(r_s,\zeta)$, for which we adopt the same analytical representation
of Ref.~\onlinecite{attacca},
\begin{equation}
  \label{ecorr} e_c(r_s,\zeta)=(e^{-\beta
    r_s}-1)e_{x}^{(6)}(r_s,\zeta)+\sum_{i=0,2} \zeta^{2i}\alpha_i(r_s),
\end{equation} where $e_{x}^{(6)}(r_s,\zeta)=
e_{x}(r_s,\zeta)-(1+\frac{3}{8}\zeta^2+\frac{3}{128}\zeta^4)e_{x}(r_s,0)$,
and the functions $\alpha_{i}(r_s)$ are defined by
\begin{eqnarray}
\label{alphaifittone} \alpha_{i}(r_s)&=&A_i+(B_i r_s+ C_i r_s^2 + D_i
r_s^3)\nonumber\\ &&\times\ln\left( 1+\frac{1}{E_i r_s +
F_i r_s^{3/2}+G_i r_s^2 +H_i r_s^3} \right).\nonumber \\ 
\end{eqnarray} We constrain the correlation energy (\ref{ecorr}) to
satisfy known high-- and low-density limits (Appendix B), reducing in
this way the number of free fitting parameters from 29 to 18. The
correlation energy of the 2V2DEG, as given by Eqs. (\ref{ecorr}) and
(\ref{alphaifittone}) with the parameters listed in
Table~\ref{param_fit}, represents a central result of this work.

\subsection{Phase diagram}

\begin{figure}[t]
\includegraphics[scale=0.8,trim=1.1cm 0 0 0]{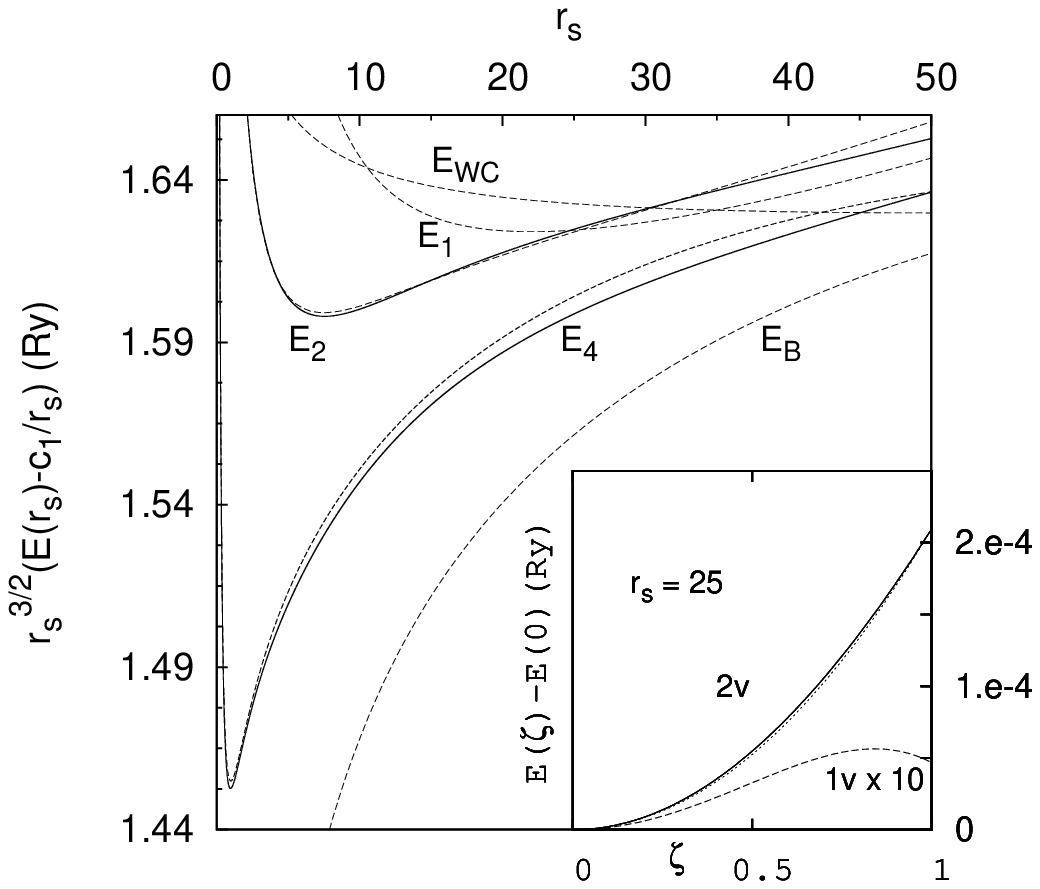}
\caption{Energy per particle of various phases of the 2DEG at $T=0$.  The energy
  label indicates the number $N_c$ of equivalent components in the homogeneous
  fluid, or the bosonic fluid (B), or the triangular Wigner crystal (WC), as
  applicable. Results of the present study are given by solid lines. The dashed
  lines are from Ref.~\onlinecite{attacca} ( $\hbox{E}_1$, $\hbox{E}_2$ ),
  Ref.~\onlinecite{conti} ($\hbox{E}_4$ ), Ref.~\onlinecite{rapisarda} (${\rm
    E}_{\rm WC}$), and Ref.~\onlinecite{depa-bosons} (${\rm E}_{\rm B}$ ). The
  inset shows $E(r_s,\zeta)-E(r_s,0)$ from Eq.~(\ref{ecorr}) (solid line)
  together with the simple quadratic dependence $[E(r_s,1)-E(r_s,0)]\zeta^2$
  (dotted), for the 2V2DEG, and the result for the 1V2DEG \cite{attacca}
  (dashed) magnified by a factor 10, at $r_s=25$. }\label{pdnostro}
\end{figure}

In Fig.~\ref{pdnostro} we plot the energies of the paramagnetic and
the fully spin-polarized 2V2DEG. They are shown by solid lines labeled
with the number of components, $N_c=4$ and $N_c=2$, respectively. For
comparison, we also plot QMC results for other phases of the 2DEG:
$N_c=1$ labels the fully polarized one-valley 2DEG \cite{attacca},
whereas the 2D charged-boson fluid  \cite{depa-bosons} corresponds to
the limit of an infinite-component 2DEG. The energy of the Wigner
crystal is known to be almost independent of the number of components
 \cite{cep89,rapisarda}; we report here the result of
Ref.~\onlinecite{rapisarda}. We note that at large $r_s$ as the number of
components increases the ground-state energy appears to quickly
approach the infinite-component limit.

The dashed line for $N_c=2$ in Fig.~\ref{pdnostro} is the result of
Ref.~\onlinecite{attacca} for the paramagnetic 2DEG. Its agreement with our
curve for the polarized 2V2DEG is expected, but still gratifying: the
two calculations differ by details in the extrapolation to the
thermodynamic limit, and the closeness of their results supports a
good control of the finite-size bias. The dashed line for $N_c=4$ is
instead the result of Ref.~\onlinecite{conti} for the paramagnetic 2V2DEG.
Its difference with the present results conveys a physical
information, namely the quantitative effect of BF correlations which
were not included in the previous simulations \cite{conti}. Backflow
improves the nodal structure of the PW wave function, yielding in the
FN approximation a tighter upper bound to the exact ground-state
energy \cite{kwon93}. It is known \cite{attacca} that BF correlations
lower the FN energy more for $N_c=2$ than for $N_c=1$. Here we find
that the BF energy gain for $N_c=4$ \cite{spin_conta} is {\em smaller}
than for $N_c=2$ (see Table \ref{back} of Appendix A), albeit 
larger than for $N_c=1$. The
modest effect of BF correlations for $N_c=4$ entails only marginal
quantitative changes to the phase diagram of the 2V2DEG predicted in
Ref.~\onlinecite{conti}. The density of Wigner crystallization shifts to a
slightly lower value, $r_s\simeq 45$.

Before discussing the spin polarization dependence of the energy and
our prediction for the spin susceptibility, we should stress that our
results provide the most accurate available estimate for the
correlation energy $e_c$ of the 2V2DEG, which in turn is the key
ingredient for density functional theory (DFT) studies of
inhomogeneous two-valley systems in 2D within the local spin density
approximation \cite{manninen-dots}.  The knowledge of $e_c$ allows
also to check the accuracy of the ansatz made in Ref.~\onlinecite{manninen}
to construct the correlation energy of a system with an arbitrary
number of components, $\epsilon_c(N_c)$, in terms of that of the one
valley system \cite{attacca}. A comparison between
$\tilde{\epsilon}_c(N_c)$ from Ref.~\onlinecite{manninen}, the present
$e_c$, and the nominally exact QMC results for charged bosons
\cite{depa-bosons} (Appendix C) exposes the limited accuracy of
$\tilde{\epsilon}_c(N_c)$ especially at large $r_s$,
including its prediction \cite{manninen} of an unphysical transition
between the nodeless ground state of the infinite-component system and
the antisymmetric ground state of the one-component 2DEG. Yet, the
comparison between DFT calculations of two-valley symmetric systems
using either $\tilde{\epsilon}_c(N_c)$ or the present $e_{c}$ would
provide a valuable test of the adequacy of $\tilde{\epsilon}_c(N_c)$
for DFT applications.

Our calculations confirm the absence of a transition from the
paramagnetic to the fully spin-polarized fluid in the two-valley
symmetric system \cite{conti}.  Moreover, in the whole density range
where the fluid  is stable we find no evidence for the stability
of a state with partial spin polarization. As illustrated in
Fig.~\ref{pdnostro} for $r_s=25$, $E(r_s,\zeta)$ displays its minimum
at $\zeta=0$ and, for all practical purposes, can be considered a
convex function of $\zeta$ \cite{hf_convex}. Convexity ensures that,
by turning on an in-plane magnetic field $B$, the absolute minimum
displayed by the  energy goes continuously from $\zeta=0$
to $\zeta=1$.  If the energy exhibits a local maximum or even an
inflection point for $\zeta <1$, instead, the $B$-driven transition to
the full spin polarization becomes a first-order one and is accompanied
by a jump in the polarization \cite{dassarma_linear}. This is clearly
the case for the 1V2DEG at $r_s=25$ also shown in Fig.~\ref{pdnostro}.

\section{Spin susceptibility}

\begin{figure}[]
  \includegraphics[scale=0.8,trim= 1.1cm 0 0 0]{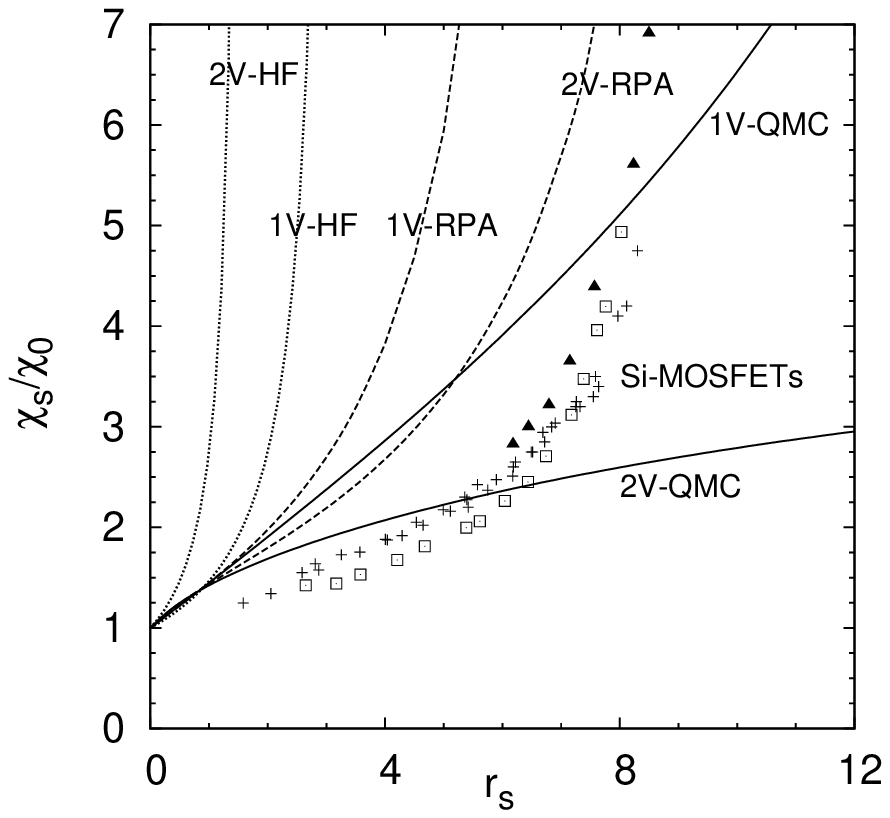}
\caption{ Spin-susceptibility enhancement of the 1V2DEG and the
2V2DEG. The results of the present work are compared with HF and
RPA \cite{dassarma,dassarma-ferrom} predictions as well as QMC results
for the one valley case \cite{attacca}. Experimental results for
Si-MOSFETs are also shown  \cite{shash_therm,pud}.}\label{chis_enh}
\end{figure}

The spin susceptibility enhancement \cite{GV} of the 2V2DEG is readily
calculated using Eq.~(\ref{ecorr}) as 
\begin{eqnarray}
\label{alphatherm-strictly2d} \chi_s/\chi_0=\Big[1-\frac{2 r_s}{\pi}+2
r_s^2\alpha_1(r_s)\Big]^{-1}.
\end{eqnarray} 
In Fig.~\ref{chis_enh} we compare our QMC prediction with the
available experimental results for electrons confined in Si-MOSFETs.
It is evident that the 2V2DEG spin susceptibility moderately
overestimates experiments at high density but largely underestimates
them at low density, where it does not display any critical growth. In
fact $\chi_s$ is a concave function of $r_s$ at all densities where
the fluid phase is stable.  Indeed, a realistic description of a 2DEG
in a solid state device requires consideration of additional elements
such as transverse thickness \cite{depalo05,dassarma} and disorder
scattering \cite{depalo05}.  As the thickness is known to suppress the
spin susceptibility and a weak disorder to enhance it, at present the
only likely candidate to explain the experimentally observed critical
behavior of $\chi_s$ \cite{shash_therm} appears to be a weak disorder.
In Fig.~\ref{chis_enh} we also report the QMC results of a 1V2DEG
\cite{attacca}. It is clear that the valley degeneracy causes a
substantial suppression of the spin susceptibility, in qualitative
agreement with the effect found in experiments on AlAs based quantum
wells \cite{gunawan}, though for an in-plane anisotropic mass.
Moreover, $\chi_s(r_s)$ changes from a convex to a concave function in
going from the 1V to the 2V2DEG.  We also show in the figure the
predictions of HF and RPA.  The general trend is that, while RPA
performs somewhat better than HF, both largely overestimate the QMC
predictions and yield divergences which either have no counterpart in
QMC, for the 2V2DEG, or in the best case  take place at a density 
about 13 times larger than in QMC, for the 1V2DEG. At least RPA reverts
the qualitatively wrong prediction of HF which yields an enhancement
of the spin susceptibility in going from the one- to the two-valley
system.

\section{Discussion and Conclusions}

We have reliably estimated the spin susceptibility of
the 2V2DEG, which provides the simplest model for electrons confined
in Si-MOSFETs. Our results clearly point to the crucial, qualitative
role of weak disorder scattering in determining the critical growth
found in the measured susceptibility  at low density \cite{shash_therm}
and to a likely minor, quantitative role of transverse thickness in
suppressing the susceptibility at high density.  
2D electron systems in high mobility Si-MOSFET's at times have been termed {\it
  clean}, meaning in fact {\it without admixture of local
  moments}\cite{shash_therm}, but also implicitly implying that observed
properties would be disorder independent and would correspond to those of an
ideally clean electron gas. This latter viewpoint, fostered
by the recent experimental observation that the effective mass enhancements of
samples\cite{mstorte,meff} with peak mobilities differing by about one order of
magnitude appear to be the same within error bars (of about 10\%), is 
contradicted by our findings.  We should
stress indeed that the samples of Refs.~\onlinecite{mstorte} and
\onlinecite{meff} are different on a number of counts and not only for the
amount of disorder.  Electrons in (111) Si-MOSFETs\cite{mstorte} have 
(i) a sizeable band mass anisotropy $m_x/m_y= 0.28$, i.e.  
comparable with the one in AlAs quantum wells\cite{gokmen}, and
(ii) a transverse thickness parameter $\sqrt(3) b/(r_s a_B)$ (see,e.g.,
Ref.~\onlinecite{depalo05} for the definition) which is more than twice the one
in (100) Si-MOSFETs\cite{meff}. Both effects (mass anisotropy and thickness) are
known to suppress spin susceptibility in an appreciable manner
\cite{gokmen,depalo05}.  Moreover, comparing the
absolute peak mobilities of Ref.~\onlinecite{mstorte} on the one hand and of
Ref.~\onlinecite{meff} on the other, i.e. of systems with quantitatively
different length and energy scales (due to different band masses) is
not appropriate. If $l$ and $a_B$ are respectively the mean-free path and the
effective Bohr radius in a given system, we find that the peak of $l/(r_sa_B) $
for the EG of Ref.~\onlinecite{mstorte} is only 3 times smaller than that
of the EG of Ref.~\onlinecite{meff}.
Hence the experiment in Ref.~\onlinecite{mstorte}
in our opinion is not at all conclusive in ruling out an effect of disorder on
the effective mass, let alone on the spin susceptibility of these systems.  

We have also
obtained: an assessment of the backflow effects on the energy of the
two-valley paramagnetic phase, which remains stable with respect to
any partially or fully polarized phase, up to the Wigner
crystallization; an analytical fit of the QMC correlation energy,
which also interpolates between exact high and low-density limits, and
provides a local spin density functional for DFT studies of two-valley
systems; the clear indication that an accurate account of correlation
beyond RPA is crucial when considering the properties of both the 1V-
and 2V2DEG.

\appendix
\section{Details of the DMC simulations}

The trial function was chosen of the usual Slater-Jastrow form,
$\Psi(R)=D(R)\,J(R)$, where $R\equiv(\mathbf{r}_1,...,\mathbf{r}_N)$
represents the coordinates of the $N$ electrons. The Jastrow factor is
a pair product, $J(R)=\exp\big[-\sum_{i<j}u(r_{ij})\big]$, with $u(r)$
the parameter-free RPA pseudopotential \cite{rapisarda}.  The phase
structure is fixed by the complex factor $D=\prod_\nu D_\nu$, i.e., a
product of Slater determinants, one for each spin-valley component.

Most of the simulations were carried out with the standard plane-wave
(PW) choice for the one-particle orbitals, $D_\nu^{{\rm PW}}={\rm
  det}[\exp(i{\mathbf k}_i\cdot {\mathbf r}_j)]$.  For $\zeta=0$ and
$\zeta=1$ we also included backflow (BF) correlations \cite{kwon93},
$D_\nu^{{\rm BF}}={\rm det}[\exp(i{\mathbf k}_i\cdot {\mathbf x}_j)]$,
where ${\mathbf x}_i={\mathbf r}_i+\sum_{j\ne i}^N
\eta(r_{ij})({\mathbf r}_i -{\mathbf r}_j)$ and the BF function
$\eta(r)$ (of the form suggested in \cite{kwon93}) was optimized by
minimization of the variational energy.  

We simulated the imaginary-time evolution of the system by a branching
random walk, using a short-time approximation of the
importance-sampled Green's function and exerting control on the number
of walkers.  Calculations were performed at
$r_s=1,\,2,\,5,\,10,\,20,\,40$.  For $\zeta=0$ and $\zeta=1$ we chose
several values of the number of electrons between $N=36$ and $N=116$,
whereas 11 intermediate values of the polarization, defined by
flipping one spin at a time, were studied for $N=52$. The twist
average, for the PW simulations, was performed on a mesh defined by
$q_x(i)=\Delta(i-1/2),\, q_y(j)=\Delta(j-1/2),\, 1\leq i \leq 8,\,
i\leq j\leq 8,\, \Delta=\pi/8L$, with $L$ the side of the simulation
box. Long-range interactions were dealt with the optimized-splitting
method of Ref.~\onlinecite{natoli-ceperley}.

Extrapolation to zero time step $\tau$ and infinite number of walkers
$N_W$ was also carried out at fixed density, on the assumption that
the $\tau$ and $N_W$ dependences are approximately independent.
Results at polarizations $\zeta=0$, $\zeta\simeq 0.5$, and $\zeta=1$
and for a bunch of $\tau$ ($N_W$) values were used to  establish the
$\tau$ ( $N_W$) dependence of the energy as function of $\zeta$; these
dependences, combined together, were then used to extrapolate to
$N_W=\infty$, $\tau=0$ the energies calculated for all values of
$\zeta$.

We record the difference between BF and PW energies at zero and full
polarization in Table \ref{back} and the whole set of energies
extrapolated to $N_W=\infty$, $\tau=0$ and including the backflow
correction in Table \ref{fit_data}.
\begin{table}
  \caption{Difference $\Delta=E^{BF}_N(r_s,\zeta)-E^{PW}_N(r_s,\zeta)$ 
    between the BF and the PW energy (in Rydberg per particle) 
    at selected values of $r_s,\, \zeta,\, N$. In parentheses 
   the statistical error on the last digit.}
\label{back}
\begin{center}
\begin{tabular}{|c|cl|cl|}
\hline
&\multicolumn{2}{|c|}{$\zeta=0$} & \multicolumn{2}{|c|}{$\zeta=1$}\\   
\hline
$r_s$ & $N$ & $\Delta$ & $N$ & $\Delta$ \\
\hline
1 & 52 & -0.0028(1)  & 50 & -0.0034(1)\\
  &    &             & 58 & -0.0035(2)\\
  &    &             & 90 & -0.0032(1)\\
\hline
2  & 52 & -0.00166(5) & 42 & -0.00175(9)\\
  &    &             & 50 & -0.00192(9)\\
  &    &             & 58 & -0.00217(9)\\
\hline
5 & 52 & -0.00057(2) & 50 & -0.00077(3)\\
  &    &             & 58 & -0.00088(3)\\
\hline
10 & 52 & -0.00021(1) & 42 & -0.00025(1)\\
   & 84 & -0.00022(1) & 50 & -0.00030(2)\\
   &    &             & 58 & -0.00032(2)\\
\hline
20 & 52 & -0.000043(6) & 42 & -0.000081(7)\\
   &    &              & 50 & -0.000085(7)\\
   &    &              & 58 & -0.000116(6)\\
   &    &              & 90 & -0.000116(6)\\
\hline
40 & 52 & -0.000020(1) & 50 & -0.000020(1)\\
   &    &              & 90 & -0.000031(1)\\
\hline
\end{tabular}
\end{center}
\end{table}
\begin{table*}[]
  \caption{Data used for the fit described in the paper. Twist-averaged 
    DMC energy in Rydberg per particle $E_N(r_s,\zeta)$, 
    calculated at finite $N$, extrapolated to zero time step and 
    infinite number of walkers, and including BF correlations; 
    in parentheses the statistical error on the last two figures shown.
    The backflow correction was obtained from Table \ref{back} employing 
    the results at the largest $N$ available.}
\label{fit_data}
\begin{center}
\begin{tabular}{|c|rcl|c|rcl|c|rcl|}
  \hline
  $r_s$& $N$ & $\zeta$ &$E_N(r_s,\zeta)$& $r_s$ & $N$&  $\zeta$ &$E_N(r_s,\zeta)$&$r_s$ & $N$&  $\zeta$ &$E_N(r_s,\zeta)$ \\
  \hline
  1 &   36 & 0      &-0.76940(15)  & 5 &   36 & 0       &-0.308540(26)  & 20&  36 & 0      &-0.0930324(80) \\  
    &   36 & 1      &-0.42501(21)  &   &   36 & 1       &-0.299849(46)  &   &  36 & 1      &-0.092705(13)  \\  
    &   52 & 0      &-0.76418(14)  &   &   52 & 0       &-0.308001(25)  &   &  42 & 1      &-0.092681(13)  \\  
    &   52 & 1/13   &-0.76192(14)  &   &   52 & 1/13    &-0.307933(26)  &   &  52 & 0      &-0.0929597(79) \\  
    &   52 & 2/13   &-0.75430(15)  &   &   52 & 2/13    &-0.307727(26)  &   &  52 & 1/13   &-0.0929559(80) \\  
    &   52 & 3/13   &-0.74537(15)  &   &   52 & 3/13    &-0.307614(27)  &   &  52 & 2/13   &-0.0929483(81) \\  
    &   52 & 4/13   &-0.73040(15)  &   &   52 & 4/13    &-0.307191(28)  &   &  52 & 3/13   &-0.0929498(83) \\  
    &   52 & 5/13   &-0.71189(16)  &   &   52 & 5/13    &-0.306660(29)  &   &  52 & 4/13   &-0.0929340(85) \\  
    &   52 & 6/13   &-0.68872(16)  &   &   52 & 6/13    &-0.305994(31)  &   &  52 & 5/13   &-0.0929046(87) \\  
    &   52 & 7/13   &-0.66258(17)  &   &   52 & 7/13    &-0.305416(33)  &   &  52 & 6/13   &-0.0928788(91) \\  
    &   52 & 8/13   &-0.63301(17)  &   &   52 & 8/13    &-0.304745(35)  &   &  52 & 7/13   &-0.0928636(96) \\  
    &   52 & 9/13   &-0.59922(18)  &   &   52 & 9/13    &-0.303896(37)  &   &  52 & 8/13   &-0.092842(10)  \\  
    &   52 & 10/13  &-0.55908(19)  &   &   52 & 10/13   &-0.302872(39)  &   &  52 & 9/13   &-0.092816(11)  \\  
    &   52 & 11/13  &-0.51827(19)  &   &   52 & 11/13   &-0.301915(42)  &   &  52 & 10/13  &-0.092765(11)  \\  
    &   52 & 1      &-0.42381(21)  &   &   52 & 1       &-0.299624(46)  &   &  52 & 11/13  &-0.092735(12)  \\  
    &   84 & 0      &-0.76258(14)  &   &   84 & 0       &-0.307778(25)  &   &  52 & 1      &-0.092659(13)  \\  
    &   84 & 1      &-0.42201(21)  &   &   84 & 1       &-0.299197(45)  &   &  84 & 0      &-0.0929138(79) \\  
    &      &        &              &   &      &         &               &   &  84 & 1      &-0.092577(13)  \\ \cline{1-8} 
$r_s$& $N$ & $\zeta$&$E_N(r_s,\zeta)$&$r_s$& $N$&  $\zeta$ &$E_N(r_s,\zeta)$& &  100& 1      &-0.092562(13)  \\  \cline{1-8}
 2  &   26 & 3/13   &-0.587078(70) & 10&   26 & 3/13    &-0.172824(16)  &   &  116& 1      &-0.092552(13)  \\ 
    &   36 & 0      &-0.590629(64) &   &   36 & 0       &-0.172782(15)  &   &     &        &               \\ \cline{9-12} 
    &   36 & 1      &-0.51883(12)  &   &   36 & 1       &-0.171014(29)  &$r_s$& $N$ & $\zeta$&$E_N(r_s,\zeta)$\\ \cline{9-12}  
    &   52 & 0      &-0.588677(63) &   &   42 & 1       &-0.171000(29)  & 40&  36 & 0      &-0.0489598(23) \\  
    &   52 & 1/13   &-0.588144(63) &   &  50  & 1       &-0.170903(29)  &   &  36 & 1      &-0.0488830(31) \\  
    &   52 & 2/13   &-0.586577(65) &   &  52  & 0       &-0.172599(15)  &   &  52 & 0      &-0.0489302(23) \\  
    &   52 & 3/13   &-0.584820(67) &   &  52  & 1/13    &-0.172555(15)  &   &  52 & 1/13   &-0.0489325(23) \\  
    &   52 & 4/13   &-0.581489(70) &   &  52  & 2/13    &-0.172524(15)  &   &  52 & 2/13   &-0.0489275(23) \\  
    &   52 & 5/13   &-0.577593(76) &   &  52  & 3/13    &-0.172522(16)  &   &  52 & 3/13   &-0.0489312(24) \\  
    &   52 & 6/13   &-0.572603(79) &   &  52  & 4/13    &-0.172425(17)  &   &  52 & 4/13   &-0.0489244(24) \\  
    &   52 & 7/13   &-0.567330(85) &   &  52  & 5/13    &-0.172309(18)  &   &  52 & 5/13   &-0.0489179(24) \\  
    &   52 & 8/13   &-0.561367(91) &   &  52  & 6/13    &-0.172163(19)  &   &  52 & 6/13   &-0.0489100(25) \\  
    &   52 & 9/13   &-0.554274(96) &   &  52  & 7/13    &-0.172056(20)  &   &  52 & 7/13   &-0.0489053(26) \\  
    &   52 & 10/13  &-0.54584(10)  &   &  52  & 8/13    &-0.171941(21)  &   &  52 & 8/13   &-0.0489042(27) \\  
    &   52 & 11/13  &-0.53737(11)  &   &  52  & 9/13    &-0.171760(23)  &   &  52 & 9/13   &-0.0488968(27) \\  
    &   52 & 1      &-0.51815(12)  &   &  52  & 10/13   &-0.171525(24)  &   &  52 & 10/13  &-0.0488833(28) \\  
    &   78 & 3/13   &-0.583684(65) &   &  52  & 11/13   &-0.171362(26)  &   &  52 & 11/13  &-0.0488779(29) \\  
    &   78 & 7/13   &-0.566625(84) &   &  52  & 1       &-0.170929(29)  &   &  52 & 1      &-0.0488635(31) \\  
    &   84 & 0      &-0.587940(61) &   &  78  & 3/13    &-0.172383(16)  &   &  78 & 7/13   &-0.0488920(26) \\  
    &   84 & 1      &-0.51703(12)  &   &  78  & 7/13    &-0.171970(20)  &   &  84 & 0      &-0.0489127(23) \\  
    &  104 & 7/13   &-0.566206(83) &   &  84  & 0       &-0.172490(15)  &   &  84 & 1      &-0.0488322(31) \\  
    &      &        &              &   &  84  & 1       &-0.170718(29)  &   &  104& 7/13   &-0.0488743(26) \\ 
    &      &        &              &   &  90  & 1       &-0.170765(29)  &   &     &        &               \\               
    &      &        &              &   &  104 & 7/13    &-0.171901(20)  &   &     &        &               \\                                                                    
\hline                                                                  
\end{tabular}
\end{center}
\end{table*}
\section{High and low density limit of the correlation energy of the four 
component 2DEG}
We directly refer to Ref.~\onlinecite{attacca} for both the
$r_s\to\infty$ limit, whose leading terms in $r_s^{-1}$ and
$r_s^{-3/2}$ are independent of the number of components \cite{cw1},
and the $r_s\to 0$ limit at $\zeta=1$, which is the same two-component
system as the one-valley case at $\zeta=0$ \cite{attacca}.

Here we only need to specify the high-density limit for the
four-component system, $\lim_{r_s\to 0} e_c(r_s,0)=A_0+B_0\, r_s \ln
r_s$. Generalizing the procedure of Ref.~\onlinecite{kimball} to the
multivalley case, we write $e_c$ as the sum of the second-order
exchange energy $e_2^{(b)}$ and the ring contribution $e_c^{(r)}$,
whose lowest order $e_2^{(r)}$ is the direct term of the second-order
energy per particle.  It turns out that $e_2^{(b)}$ is a constant,
independent of $r_s$ and the number of components of the system,
$N_c$, while $e_2^{(r)}=e_2^{(r)}(N_c)$ depends only on $N_c$.
Furthermore, we notice that the ring contribution scales with $N_c$ as
$e_c^{(r)}(r_s,N_c)=N_c\,f(r_s\,N_c^{3/2})$, so that the following
scaling law holds
\begin{eqnarray}
\label{scaling}
e_c^{(r)}(r_s,4)=2e_c^{(r)}(4\sqrt{2}r_s,2).
\end{eqnarray}
By applying the scaling law (\ref{scaling}) to the leading terms of
$e_c$, we find $A_0 =e^{(b)}_2+ 2\,e^{(r)}_2(2)=-0.99870$, $B_0 =
16\,(10-3 \pi)/(3 \pi)$.
\section{Check of  an approximate multicomponent correlation energy}
\begin{figure}
  \includegraphics[scale=0.85,trim = 1.2cm 0 0 0]{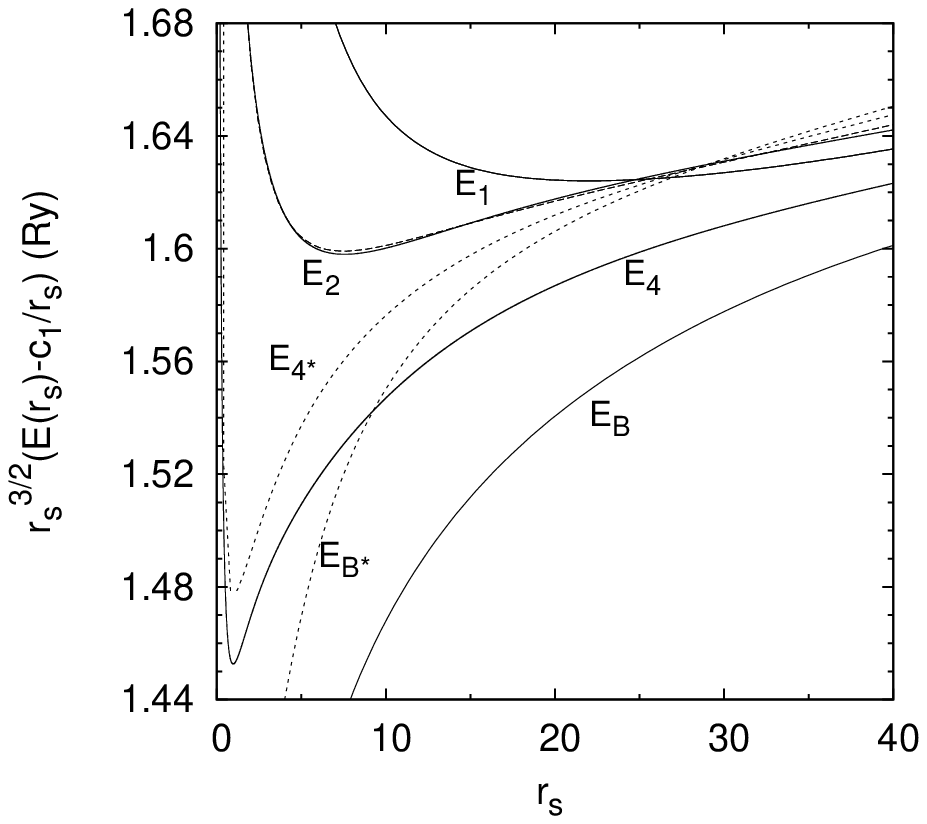}
  \caption{Phase diagram of the multicomponent 2DEG: liquid phases.
    Solid $\hbox{E}_4$ and $\hbox{E}_2$ from the present work, solid
    $\hbox{E}_1$ and long-dashed $\hbox{E}_2$ from
    Ref.~\onlinecite{attacca}, ${\rm E}_{\rm B}$
    from Ref.~\onlinecite{depa-bosons}.  $\hbox{E}_{4^*}$ and
    ${\rm E}_{\rm B^*}$ are from
    Ref.~\onlinecite{manninen}. One- and two-component energies from
    Ref.~\onlinecite{manninen} coincide with
    Ref.~\onlinecite{attacca}'s ones by construction.  }\label{pdmann}
\end{figure}

In Fig.~\ref{pdmann} we show a comparison between the multicomponent
correlation energy $\tilde{\epsilon}_c(N_c)$ of Ref.~\onlinecite{manninen} and
various simulation results, including the present two-valley
calculation, and the nominally exact QMC results for charged
bosons \cite{depa-bosons}. Total energies are displayed. The scale of
the figure, emphasizes 
the limited accuracy of $\tilde{\epsilon}_c(N_c)$ in the large $r_s$ regime.


\begin{thebibliography}{99}
\bibitem{note-chi}Indeed,
for the one-valley  2DEG  the RPA grossly overestimates
$\chi_s$ \cite{depalo06} and    predicts  a divergence at
moderate low-density\cite{dassarma-ferrom} which can only be
ascribed to its limited accuracy. Similarly, a small but  unjustified  
change  of the correlation energy in the
calculations carried out for the 2V2DEG within the classical mapping 
changes a $\chi_s$ finite and   smooth at all densities in a diverging one.
\bibitem{splitting}Evidently, we  disregard the valley splitting
predicted when going beyond the effective mass
approximation \cite{ando}.
\bibitem{hf_convex} Actually,  the convexity of $E(r_s,\zeta)$ is spoiled by the
  exchange term at some finite polarization $\zeta_c$. However, for
  $r_s\agt 1$, correlation
  effects push $\zeta_c$ exponentially close to   1, in a range 
$1-\zeta_c \propto e^{-\beta r_s}$.
\bibitem{cw1}The coefficients of the leading powers of $r_s$ are kept
fixed to their analytical values for the crystal phase in
Ref.~\onlinecite{attacca}, whereas they are fitting parameters in
the present work.
\end{thebibliography}
\end{document}